\documentclass[
    ,final            
  ]
  {aipproc}
\layoutstyle{8x11single}

\RequirePackage{xspace}
\def\Vub  {\ensuremath{|V_{ub}|}\xspace}
\def\Vcb  {\ensuremath{|V_{cb}|}\xspace}
\def\Vts  {\ensuremath{|V_{ts}|}\xspace}
\def\Vtd  {\ensuremath{|V_{td}|}\xspace}
\def\B    {\ensuremath{B}\xspace}
\def\Bz   {\ensuremath{B^0}\xspace}
\def\Bu   {\ensuremath{B^+}\xspace}
\def\CP   {\ensuremath{C\!P}\xspace}
\def\BF   {$B$ Factory}
\usepackage{relsize}
\def\babar{\mbox{\slshape B\kern-0.1em{\smaller A}\kern-0.1em
    B\kern-0.1em{\smaller A\kern-0.2em R}}}

\begin{document}
\vbox{\hfill{CLNS 03/1855}}
\title{Recent CLEO CKM Results\footnote{
The following article has been submitted to the Proceedings of Beauty 2003.
After it is published, it will be found at
\protect\url{http://proceedings.aip.org}.}}

\author{Karl M.~Ecklund
\\ {\normalsize\sl (Representing the CLEO Collaboration)}
}{
address={Laboratory of Elementary-Particle Physics, Cornell University
         Ithaca, NY 14853, U.S.A.}
}

\begin{abstract}
I report $B$ physics results from the CLEO collaboration, highlighting
measurements of the Cabibbo-Kobayashi-Maskawa matrix elements \Vcb and
\Vub.   I report a recent measurement of \Vub through study of the
$q^2$ dependence of 
$\bar{B}\to\pi\ell\bar{\nu}$ and $\bar{B}\to\rho\ell\bar{\nu}$.
I also describe new measurements of the inclusive
semileptonic branching fraction ${\cal B}(\bar{B}\to X e \bar{\nu})$
and of moments of the hadronic invariant mass spectrum in 
$\bar{B}\to X \ell \bar{\nu}$, with impact on \Vcb.
\end{abstract}

\maketitle


\section{Introduction}
CLEO's recent measurements of the Cabibbo-Kobayashi-Maskawa matrix
elements \Vub and \Vcb are still competitive in the era of \BF\ 
statistics because these measurements are systematically and
theoretically limited, and CLEO's well-understood detector and
analysis techniques bring added value to the world knowledge of these
couplings in semileptonic \B decays.  Along with \Vtd and \Vts
inferred from measurements of \B mixing,  \Vub and \Vcb form an
important part of the \B \CP puzzle: the sides of the Unitarity Triangle.
Combined with measurements of the angles from the \CP-violating
phases, will the Unitarity Triangle hold together or indicate the
presence of new physics?  Precise measurements of \Vub and \Vcb are
required for this test.

\newsavebox{\fool}
\sbox{\fool}{\raisebox{1.2pt}{$\mathbf\Vub$}}
\section{Exclusive \usebox{\fool} Measurement}
\label{sec:vub}
\begin{figure*}
\resizebox{!}{5.75cm}{\includegraphics{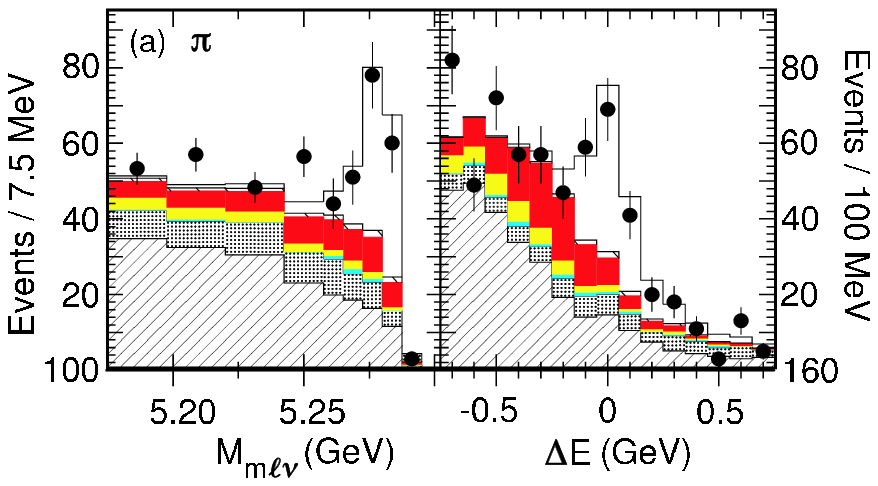}}
\resizebox{!}{6.25cm}{\includegraphics{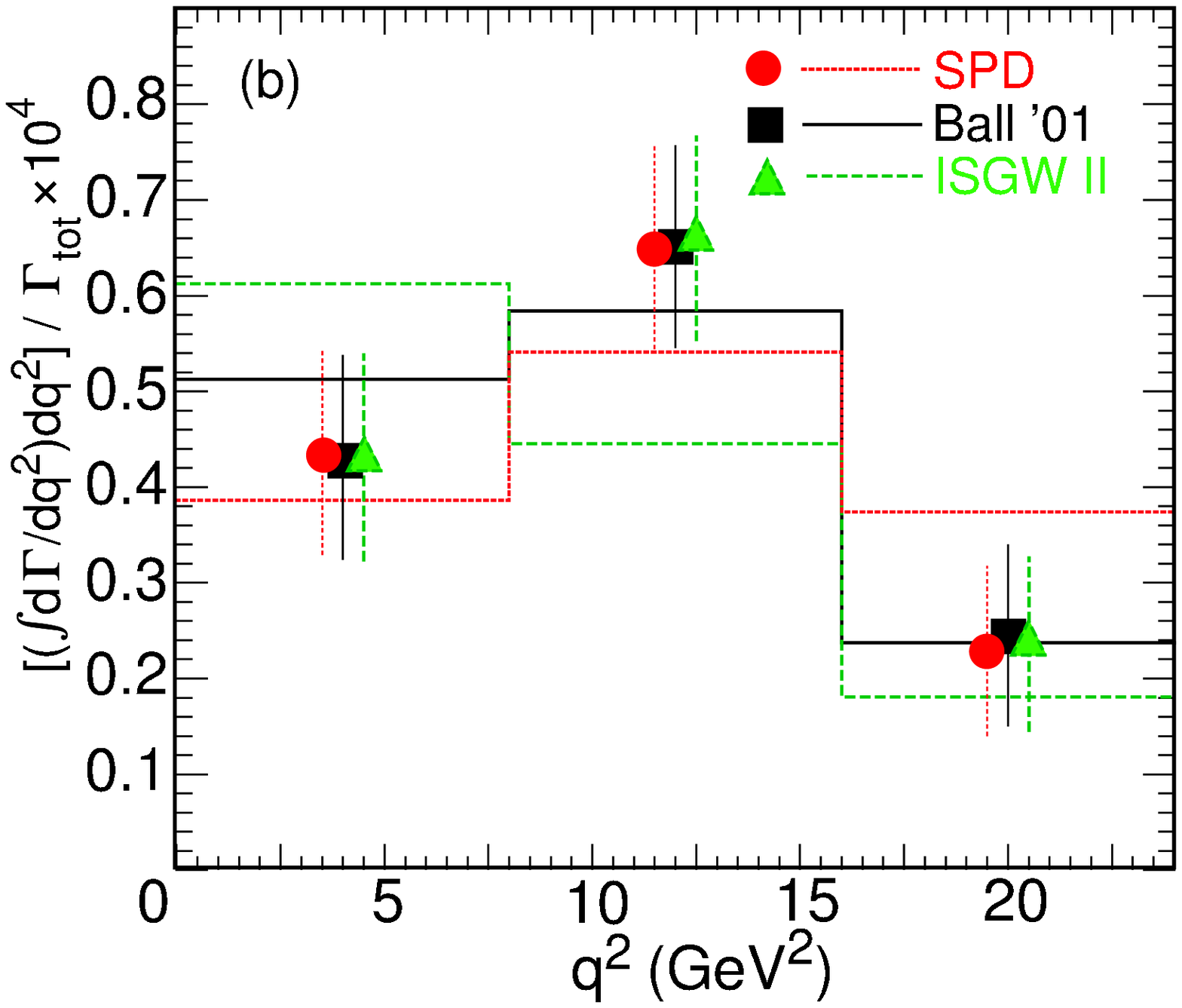}}
\caption{Exclusive $B \to \pi\ell\bar{\nu}$: (a) projections of ML fit to
$M_{m\ell\nu}$ and $\Delta E$ and (b) fit to $d\Gamma/dq^2$.}
\label{kme:fig.ExcVub}
\end{figure*}

Recently CLEO measured \Vub in the exclusive modes 
${\bar B}\to [\pi/\rho/\omega/\eta]\ell\bar{\nu}$ \cite{Athar:2003yg}.
The neutrino is reconstructed from the missing energy and
momentum of the event, taking advantage of CLEO's large solid angle
(95\%).  Tracks reconstructed from hits in the drift chamber and
silicon are combined with neutral showers in the calorimeter to form
missing energy and momentum estimates.  Considerable effort is made to
remove spurious tracks and showers from hadronic interactions, in
order to give the best estimate of the neutrino energy and momentum.
When the neutrino candidate is combined with a lepton and
light meson candidate, energy and momentum conservation leads to
signal peaks in $\Delta E = E - E_{\rm beam}$ and the $B$ candidate
 invariant mass $M_{m\ell\nu}$, with $S/B \approx 1$.  We perform a
simultaneous maximum likelihood fit in 
$\Delta E$ and $M_{m\ell\nu}$ to seven sub-modes: 
$\pi^\pm$, $\pi^0$, $\rho^\pm$, $\rho^0$, 
$\omega/\eta \to \pi^+\pi^-\pi^0$, and $\eta \to \gamma\gamma$.  
In the fit we use isospin symmetry to constrain the semileptonic widths
$\Gamma^{SL}({\pi^\pm})=2\Gamma^{SL}({\pi^0})$ and 
$\Gamma^{SL}({\rho^\pm})=2\Gamma^{SL}({\rho^0})\approx 2\Gamma^{SL}({\omega})$,
where the final approximate equality is inspired by constituent quark
symmetry.  We find clear signals for $\pi$ and $\rho/\omega$ and a 3.2
sigma significance for $\eta\ell\bar{\nu}$.  The branching fractions are
given in Table~\ref{kme:tab.b2ubf}.

\begin{table}
\caption{Branching fractions for exclusive $b\to u \ell\bar{\nu}$ modes.
In order, the errors given are statistical, systematic, due to
the $\pi$ form factor, and due to $\rho$ form factors.  For
$\eta\ell\bar{\nu}$, the last error is due to the $\eta$ form factor.}
\label{kme:tab.b2ubf}
\begin{tabular}{cc}
Mode & BF $\times 10^4$ \\ \hline
$B^0\to \pi^-\ell^+\nu$  
	& $1.33 \pm 0.18 \pm 0.11 \pm 0.01 \pm 0.07$ \\
$B^0\to \rho^-\ell^+\nu$ 
	& $2.17 \pm 0.34\ ^{+0.47}_{-0.54} \pm 0.01 \pm 0.41 $ \\
$B^+\to \eta\ell^+\nu$   
	& $0.84 \pm 0.31 \pm 0.10 \pm 0.09 $
\end{tabular}
\end{table}

Signals for $\pi$ (Fig.~\ref{kme:fig.ExcVub}a) and $\rho$ are extracted
separately in three $q^2$ bins.  The differential decay rate for 
$\pi\ell\bar{\nu}$, 
\begin{displaymath}
{d\Gamma \over dq^2} = {G_F^2 \over 24 \pi^3} \Vub^2 p_\pi^3 
\left|f_+(q^2)\right|^2,
\end{displaymath}
includes a form factor $f_+$ which encodes the hadronic physics for the
$B\to\pi$ transition.  For $\rho\ell\bar{\nu}$, which has a vector meson in
the final state, there are two additional form factors.
Given form factors from theory, we
extract \Vub from a fit to $d\Gamma/dq^2$ (Fig.~\ref{kme:fig.ExcVub}b). 
Combining $\bar{B}\to \pi\ell\bar{\nu}$ and
$\bar{B}\to\rho\ell\bar{\nu}$ results we find 
\begin{displaymath}
|V_{ub}| =  
(3.17 \pm 0.17           |_{\rm st}
      \ ^{+0.16}_{-0.17} |_{\rm sy}
      \ ^{+0.53}_{-0.39} |_{\rm th}
      \pm 0.03           |_{\rm ff}   ) \times 10^{-3},
\end{displaymath}
where the quoted errors are, in order, statistical, experimental
systematic, and theoretical from form factor normalization and shape.
This result uses form factors from Lattice QCD ($q^2 > 16$ GeV$^2$) and
light cone sum rules ($q^2 > 16$ GeV$^2$) where each are most reliable.
In a test of $\bar{B}\to\pi\ell\bar{\nu}$ form factors, ISGW2
\cite{Scora:1995ty} is disfavored (Fig.~\ref{kme:fig.ExcVub}b).

By binning in $q^2$, this analysis has relaxed the theoretical
assumption on the shape of the form factor made in earlier
analyses. The theoretical uncertainty on the form factor normalization
currently limits the precision of the \Vub extraction.  In the future,
unquenched Lattice QCD calculations \cite{El-Khadra:b2k3} can improve the
$\bar{B}\to\pi\ell\bar{\nu}$ form factor in a limited region of
$q^2$. We find good agreement between measurements of \Vub using
inclusive techniques
\cite{Bornheim:2002du,Aubert:2003zw,Kakuno:2003fk} and other exclusive
measurements \cite{Behrens:1999vv,Aubert:2003zd}.

\sbox{\fool}{\raisebox{1.2pt}{$\mathbf\Vcb$}}
\section{Inclusive \usebox{\fool} Measurement}
A measurement of \Vcb is possible using the inclusive semileptonic
decay rate.  The experimental inputs are the branching 
fraction for $\bar{B}\to X_c\ell\bar{\nu}$ and the $B$ lifetime.  The
inclusive decay rate $\Gamma_c^{SL} = \gamma_c \Vcb^2$, where
$\gamma_c$ comes from theory.

Within the framework of heavy quark effective theory (HQET)
\cite{ManoharWise:2000dt}, the inclusive semileptonic
decay rate is expanded in a double series in $\alpha_s^n$ and
$1/M^n$, where $M$ is the heavy quark mass.  Hadronic effects enter
both in the perturbative expansion and as expansion parameters,
defined to be matrix elements of non-perturbative QCD operators.  At
${\cal O}(1/M^2)$ there are two parameters: $\lambda_1$, which is
proportional to the kinetic energy of the $b$ quark in the $B$ meson,
and $\lambda_2$, which comes from the chromomagnetic operator.  An
additional parameter $\bar\Lambda$ relates the $B$ meson mass to the $b$
quark mass. From the $B$-$B^*$ mass difference, $\lambda_2=0.128\pm0.010$
GeV$^2$. The other parameters can be estimated (\textit{e.g.}~in quark
models) but they can also be measured using spectral moments in
inclusive $B$ decay. Moments, \textit{e.g.}~of the lepton energy
spectrum, are also computed in HQET, allowing extraction of $\lambda_1$
and $\bar\Lambda$ from two or more spectral measurements.

\subsection{Inclusive Semileptonic Branching Fraction}

\begin{figure}
\resizebox{7cm}{!}{\includegraphics{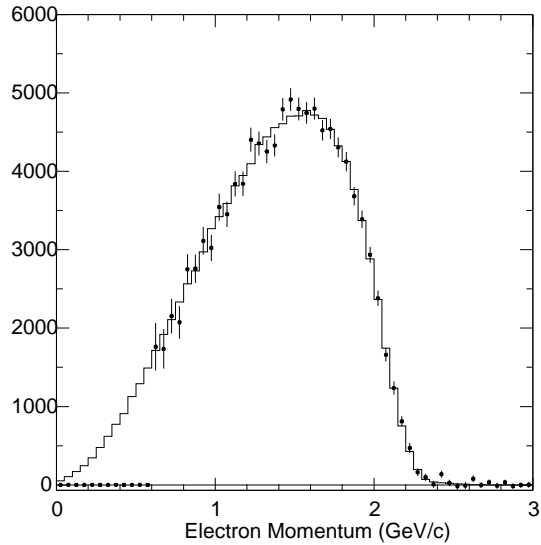}}
\caption{Unfolded primary $\bar{B}\to X e\bar{\nu}$ electron spectrum
measured using a high momentum lepton tag. The line shows a fit to exclusive
$\bar{B}\to X e\bar{\nu}$ decays, used to extrapolate below the cut at 600
MeV/$c$. (Preliminary)}
\label{kme:fig.b2xenu}
\end{figure}

CLEO has a new preliminary measurement of the inclusive semileptonic
branching fraction using a high-momentum ($p>1.5$ GeV/$c$) lepton tag.
The analysis is an update of Ref.~\cite{Barish:1996cx}, where the
lepton tag  identifies a sample of $B$ decays with high purity (98\%).
Additional electrons may come from the decay chain of the same $B$ or
from the decay of the other $B$ meson in the event 
($e^+e^-\rightarrow\Upsilon(4S)\rightarrow B\bar{B}$). 
Secondary electrons ($b\to c\to e$) and primary electrons are
separated using kinematic and charge correlations, with a known
correction from $B^0$-$\bar{B^0}$ mixing.  The new semileptonic
branching fraction is  $(10.88\pm0.08\pm0.33)$\%, in agreement with
measurements from LEP and \BF\ data \cite{HFAG:2003s}.  The spectrum
of electrons above 600 MeV/$c$ is also obtained
(Fig.~\ref{kme:fig.b2xenu}), from which spectral moments will be measured.

\subsection{Extraction of \Vcb using the Heavy Quark Expansion}
Using CLEO's new inclusive branching fraction of $(10.8\pm 0.3)$\%,
subtracting a 1\% relative contribution from $b\to u\ell\bar{\nu}$, and the
PDG average \Bz and \Bu lifetimes \cite{pdg03}, we find the
semileptonic decay rate 
$\Gamma^{SL}_c=(0.44 \pm 0.02)\times 10^{-10}$ MeV.
Using the HQET expansion for the decay rate and HQET parameters
$\bar\Lambda$ and $\lambda_1$ from CLEO measurements of the moments of
the $B\to X_s\gamma$ photon spectrum \cite{Chen:2001fj} and $\bar{B}\to
X\ell\bar{\nu}$ hadronic mass spectrum \cite{Cronin-Hennessy:2001fk}, we
obtain  $\Vcb=0.0411 \pm 0.0005|_{\bar\Lambda,\lambda_1} 
                     \pm 0.0007|_\Gamma \pm 0.0009|_{HQE}$.
The overall precision of 3\% is limited by theoretical uncertainties
from the unknown ${\cal O}(1/M^3)$ heavy quark expansion parameters.
There is an unquantifiable error from the parton-hadron duality
assumption of the inclusive approach.

\begin{figure*}
\resizebox{0.25\textwidth}{!}{\includegraphics{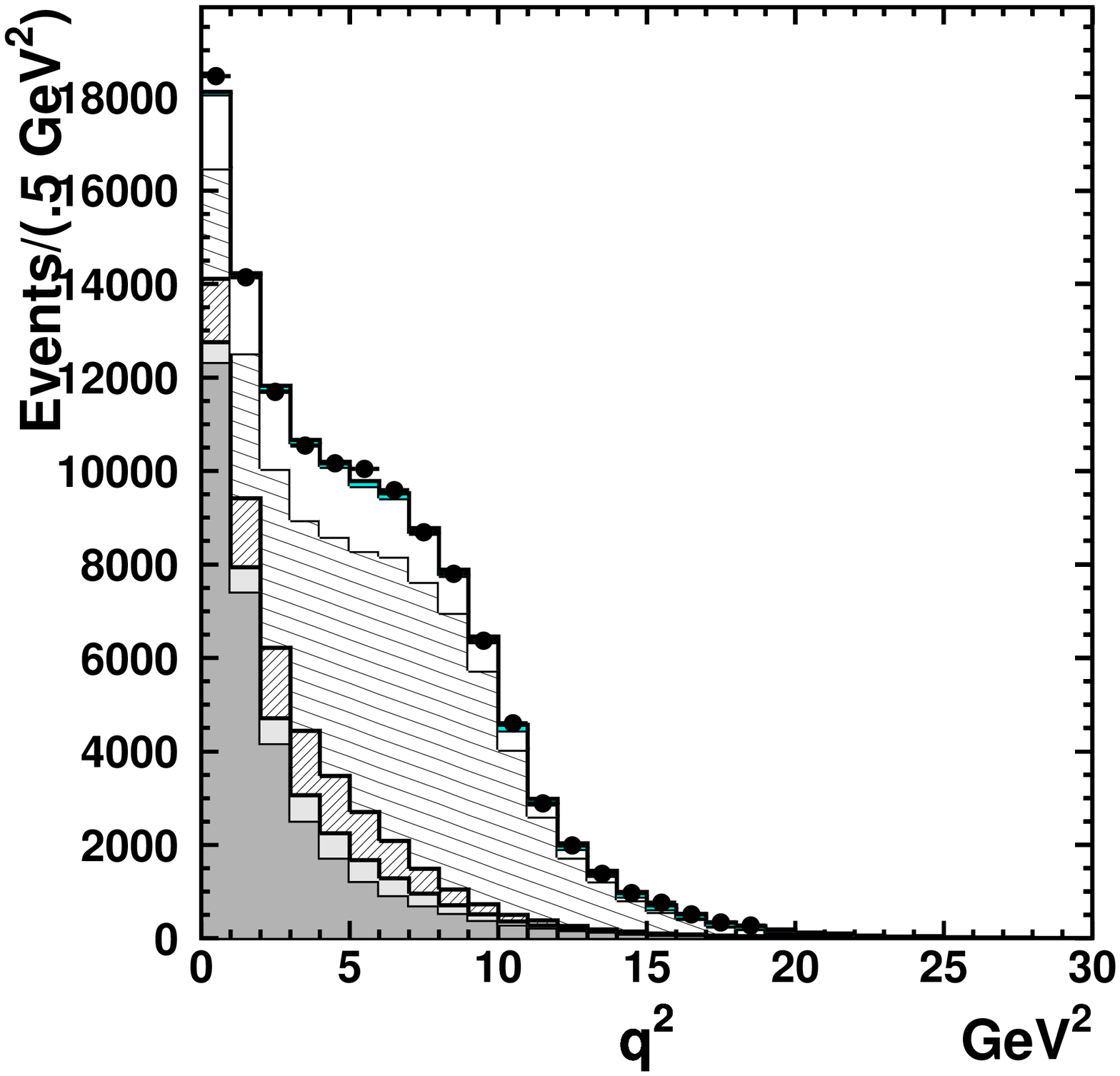}}
\resizebox{0.25\textwidth}{!}{\includegraphics{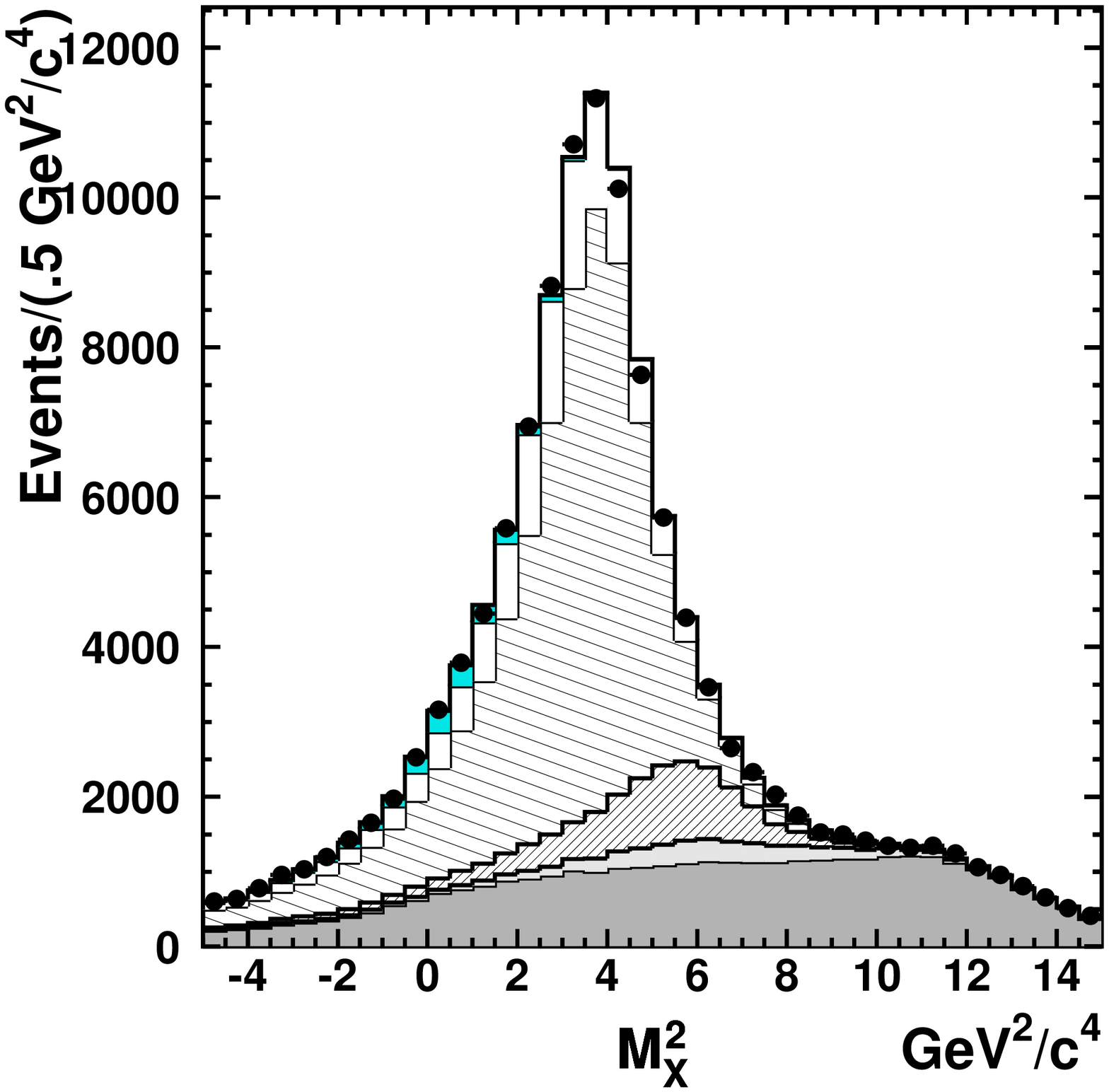}}
\resizebox{0.25\textwidth}{!}{\includegraphics{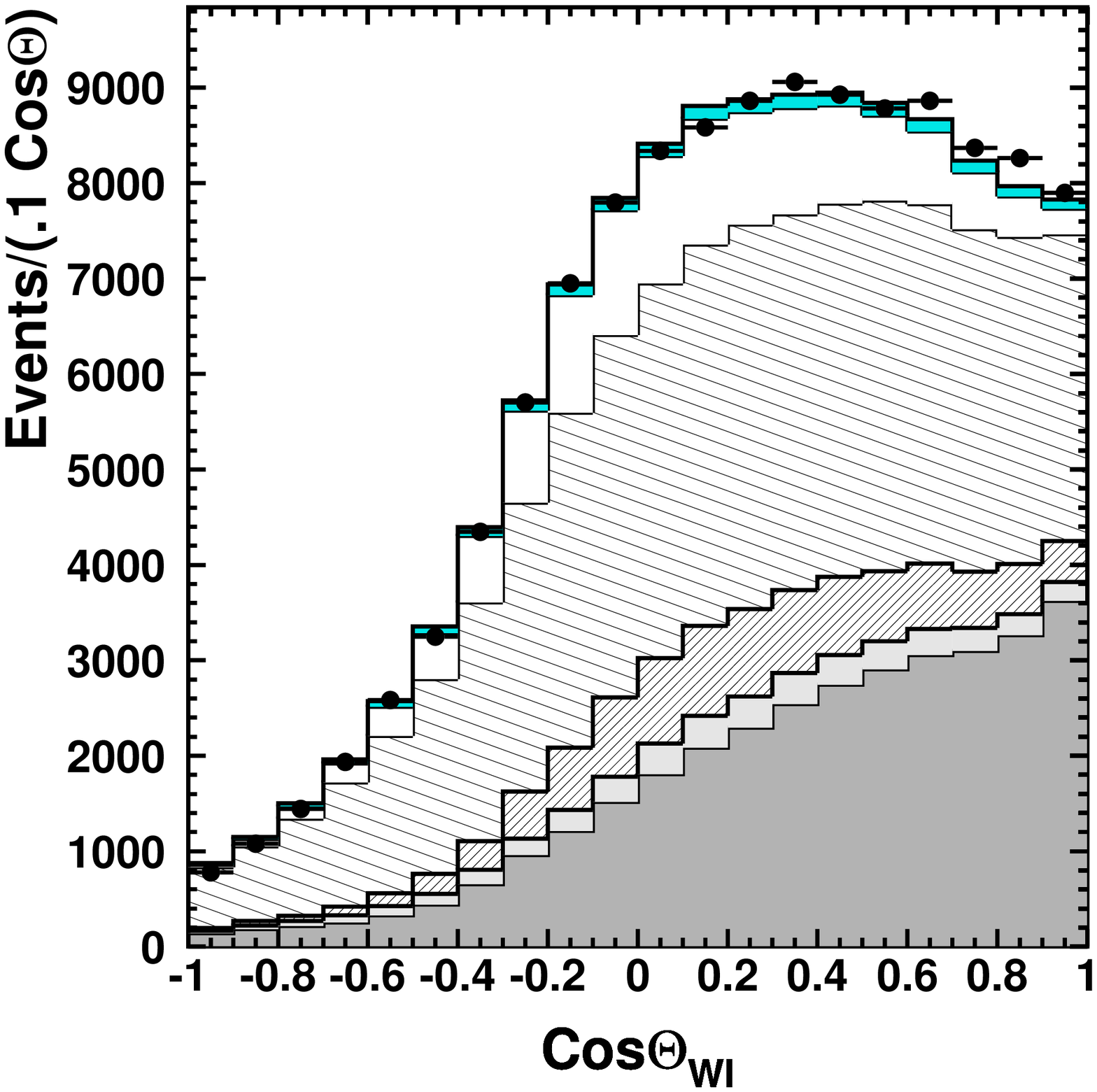}}
\resizebox{0.25\textwidth}{!}{\includegraphics{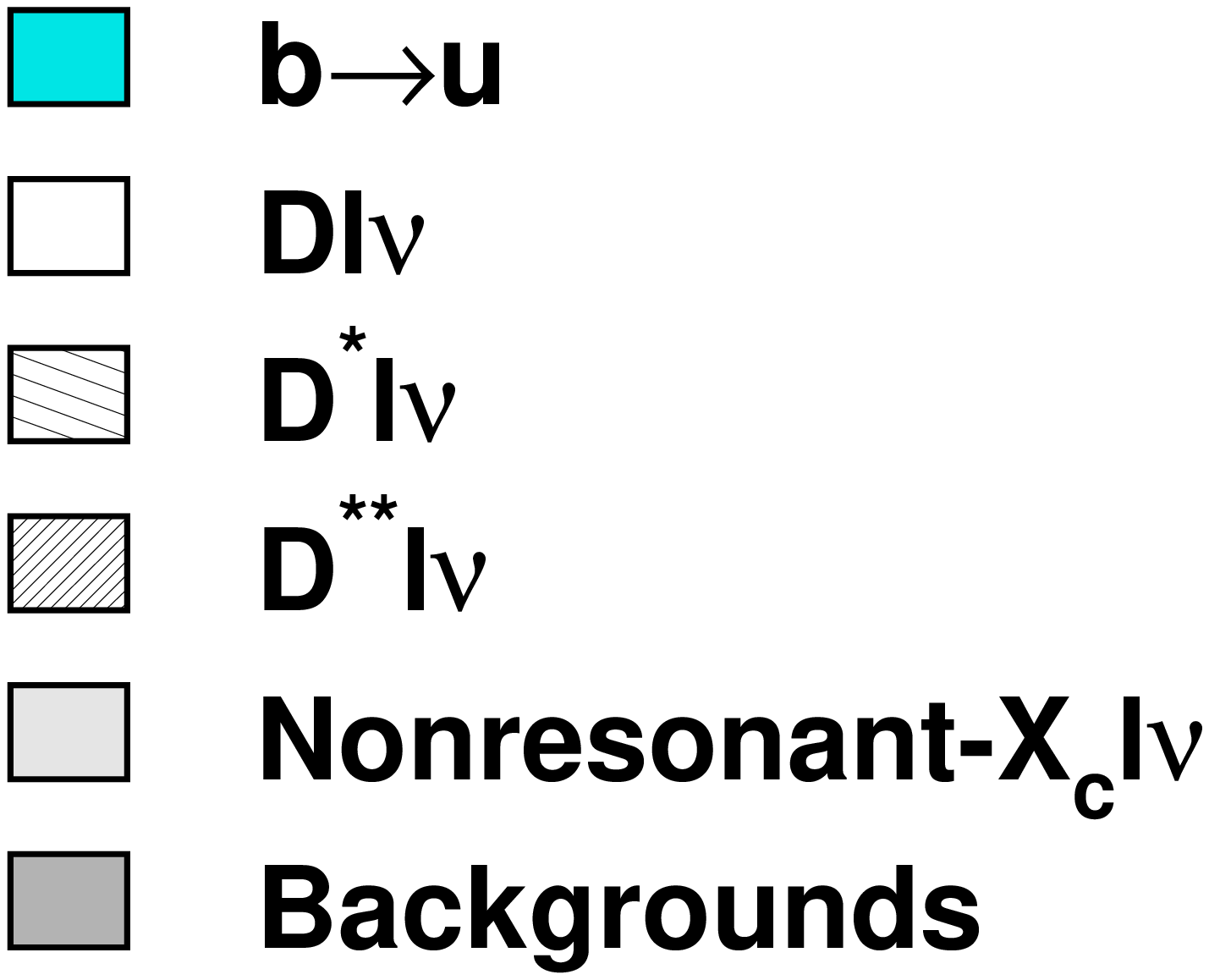}}
\caption{Projections of likelihood fit to differential decay rate.
For the three independent kinematic variables we have chosen $q^2$,
$M_X^2$, and $\cos\theta_{W\ell}$, the helicity angle of the virtual
$W$ decay.}
\label{kme:fig.b2clnu}
\end{figure*}

\subsection{Spectral Moments as a test of HQE}
It is essential to test the heavy quark expansion and the assumption
of parton-hadron duality implicit in inclusive determinations of \Vcb.
Previously CLEO published an analysis of the lepton energy spectral
moments in $\bar{B}\to X_c\ell\bar{\nu}$ with a cut at 1.5 GeV
\cite{Mahmood:2002tt}, showing good agreement with HQET expectations
and the hadronic mass and $B\to X_s\gamma$ photon spectrum analyses.
At Lepton Photon 2003, CLEO presented new preliminary results from an
analysis of the hadronic invariant mass spectrum in 
$\bar{B}\to X_c\ell\bar{\nu}$ 
\cite{Huang:2003ay}.

Like the exclusive \Vub measurement, we use the neutrino
reconstruction technique to estimate the neutrino energy and
momentum.  Combined with a lepton (electron or muon) candidate with
$p>1.0$ GeV$/c$ we can reconstruct the hadronic invariant mass
recoiling against the lepton and neutrino:
\begin{displaymath}
M_X^2 = M_B^2 + q^2 - 2 E_{\textrm{beam}} (E_\ell+ E_\nu) 
                    + 2 |\vec p_B| |\vec q| cos \theta_{Bq},
\end{displaymath}
where $q^2 ({\vec q})$ is the lepton-neutrino (virtual $W$) invariant
mass squared (momentum). 
Only the last term in the exact equation for $M_X^2$ is unknown;
fortunately because it is small in the $\Upsilon(4S)$ rest frame, the
approximation from neglecting this term is adequate.

\begin{figure}
\resizebox{10cm}{!}{\includegraphics{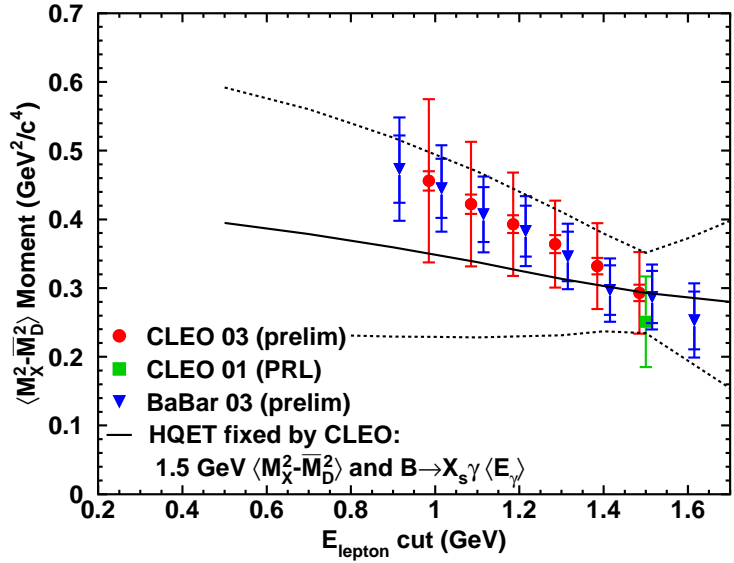}}
\caption{Hadronic recoil mass moments versus the lepton energy cut.
 For both CLEO and \babar, the inner (outer) error bars give the
 statistical (total) uncertainty.  N.B., there is substantial correlation
 among the points as the lepton energy cut is varied.  The CLEO 01
 point corresponds to \cite{Cronin-Hennessy:2001fk}.}
\label{kme:fig.mx2}
\end{figure}

We fit the three-dimensional differential decay rate to contributions
from $\bar{B}\to D\ell\bar{\nu}$, $\bar{B}\to D^*\ell\bar{\nu}$,
$\bar{B}\to D^{**}\ell\bar{\nu}$, $\bar{B}\to X_c\ell\bar{\nu}$, and
$\bar{B}\to X_u\ell\bar{\nu}$ (Fig.~\ref{kme:fig.b2clnu}), 
allowing contributions from fake lepton, $e^+e^-\to q\bar{q}$, and
$b\to c\to \ell$ backgrounds, which are estimated using data and Monte
Carlo simulation. 
From the fit results we extract the first and second moments of the
hadronic recoil mass.  The first moment 
$\left\langle M_X^2-{\bar M^2_D}\right\rangle$
is computed as a function of the minimum lepton energy cut
$E_\ell^{\textrm{min}}$, varied between 1.0 and 1.5 GeV
(Table~\ref{kme:tab.mx2}).
The first moment is plotted versus the lepton energy cut in
Fig.~\ref{kme:fig.mx2}, overlayed with results from a similar analysis
of \babar~\cite{langenegger} and expectations from HQET given the CLEO
measurement of $\left\langle E_\gamma \right\rangle$ in $B\to
X_s\gamma$.  The dotted lines show the range of uncertainty from
$1/M^3$ terms in the heavy quark expansion.  

There is good agreement with theory and between experiments, which use
complementary techniques.  \babar\ uses a reconstructed \B tag with
smaller systematic but larger statistical uncertainty, while without
the \B tag, CLEO has higher efficiency but larger backgrounds and
attendant systematic uncertainties, notably from $e^+ e^-\to q\bar{q}$.

\begin{table}
\caption{Hadronic recoil mass moments versus the lepton energy
cut. The errors on the entries in the table are the statistical,
detector systematics, and model dependence, respectively.} 
\label{kme:tab.mx2}
\smallskip
\begin{tabular}{cc}
$E_\ell^{\textrm{min}}$ Cut (GeV)& $\left\langle M_X^2-{\bar M^2_D}\right\rangle$ (GeV$^2/c^4$)\\
\hline
 1.0 &~ 0.456 $\pm$ 0.014 $\pm$ 0.045 $\pm$ 0.109 \\ 
 1.1 &~ 0.422 $\pm$ 0.014 $\pm$ 0.031 $\pm$ 0.084 \\ 
 1.2 &~ 0.393 $\pm$ 0.013 $\pm$ 0.027 $\pm$ 0.069 \\ 
 1.3 &~ 0.364 $\pm$ 0.013 $\pm$ 0.030 $\pm$ 0.054 \\ 
 1.4 &~ 0.332 $\pm$ 0.012 $\pm$ 0.027 $\pm$ 0.055 \\ 
 1.5 &~ 0.293 $\pm$ 0.012 $\pm$ 0.033 $\pm$ 0.048 
\end{tabular}
\end{table}

\section{Conclusion}
CLEO's direct contributions to beauty physics are nearing an end, but
recent measurements of CKM matrix elements \Vcb and \Vub are still
among the best available.  In the future CLEO's contribution to
flavor physics and the test of the CKM paradigm for \CP-violation will
come from measurements at charm threshold \cite{Asner:b2k3}.




\bibliographystyle{aipprocl} 


\end{document}